\documentclass[sn-mathphys]{sn-jnl}% Math and Physical Sciences Reference Style
\jyear{2021}%

\theoremstyle{thmstyleone}%
\theoremstyle{thmstyletwo}%

\theoremstyle{thmstylethree}%

\begin{document}

\title[LTSP: Long-Term Slice Propagation for Accurate Airway Segmentation]{LTSP: Long-Term Slice Propagation for Accurate Airway Segmentation}

%%=============================================================%%
%% Prefix	-> \pfx{Dr}
%% GivenName	-> \fnm{Joergen W.}
%% Particle	-> \spfx{van der} -> surname prefix
%% FamilyName	-> \sur{Ploeg}
%% Suffix	-> \sfx{IV}
%% NatureName	-> \tanm{Poet Laureate} -> Title after name
%% Degrees	-> \dgr{MSc, PhD}
%% \author*[1,2]{\pfx{Dr} \fnm{Joergen W.} \spfx{van der} \sur{Ploeg} \sfx{IV} \tanm{Poet Laureate} 
%%                 \dgr{MSc, PhD}}\email{iauthor@gmail.com}
%%=============================================================%%

\author[1,2]{\fnm{Yangqian} \sur{Wu}}\email{wyq19981114@sjtu.edu.cn}

\author[1,2]{\fnm{Minghui} \sur{Zhang}}\email{minghuizhang@sjtu.edu.cn}

\author[1,2]{\fnm{Weihao} \sur{Yu}}\email{yuweihao@sjtu.edu.cn}

\author[1,2]{\fnm{Hao} \sur{Zheng}}\email{zhenghaobs@sjtu.edu.cn}

\author[1,2]{\fnm{Jiasheng} \sur{Xu}}\email{xujiasheng@sjtu.edu.cn}

\author*[1,2]{\fnm{Yun} \sur{Gu}}\email{	geron762@sjtu.edu.cn}

\affil[1]{\orgdiv{Institute of Medical Robotics}, \orgname{Shanghai Jiao Tong University}, \orgaddress{\city{Shanghai},  \country{China}}}

\affil[2]{\orgdiv{Institute of Image Processing and Pattern Recognition}, \orgname{Shanghai Jiao Tong University}, \orgaddress{\city{Shanghai},  \country{China}}}

% \affil[3]{\orgdiv{Department}, \orgname{Organization}, \orgaddress{\street{Street}, \city{City}, \postcode{610101}, \state{State}, \country{Country}}}

\abstract{

\textbf{Purpose:} Bronchoscopic intervention is a widely-used clinical technique for pulmonary diseases, which requires an accurate and topological complete airway map for its localization and guidance. The airway map could be extracted from chest computed tomography (CT) scans automatically by airway segmentation methods. Due to the complex tree-like structure of the airway, preserving its topology completeness while maintaining the segmentation accuracy is a challenging task.\qquad\qquad\qquad

\textbf{Methods:} In this paper, a long-term slice propagation (LTSP) method is proposed for accurate airway segmentation from pathological CT scans. We also design a two-stage end-to-end segmentation framework utilizing the LTSP method in the decoding process. Stage 1 is used to generate a coarse feature map by an encoder-decoder architecture. Stage 2 is to adopt the proposed LTSP method for exploiting the continuity information and enhancing the weak airway features in the coarse feature map. The final segmentation result is predicted from the refined feature map.

\textbf{Results:} Extensive experiments were conducted to evaluate the performance of the proposed method on 70 clinical CT scans. The results demonstrate the considerable improvements of the proposed method compared to some state-of-the-art methods as most breakages are eliminated and more tiny bronchi are detected. The ablation studies further confirm the effectiveness of the constituents of the proposed method.\qquad\qquad

\textbf{Conclusion:} Slice continuity information is beneficial to accurate airway segmentation. Furthermore, by propagating the long-term slice feature, the airway topology connectivity is preserved with overall segmentation accuracy maintained. \qquad\qquad\qquad\qquad\qquad\qquad\qquad\qquad\qquad\qquad\qquad\qquad

}


%%\pacs[JEL Classification]{D8, H51}

%%\pacs[MSC Classification]{35A01, 65L10, 65L12, 65L20, 65L70}

\maketitle

\section{Introduction}\label{sec1}
Bronchoscopic intervention \cite{higgins2015multimodal, mehta2018evolutional, shen2019context} is widely used in clinical practice since it could achieve minimally invasive access techniques for pulmonary diseases. To improve the efficacy of the bronchoscopic intervention, an accurate 3D airway map extracted from chest computed tomography (CT) is required, which is essential to the endoscopic tips localization and intraoperative guidance. However, due to the complex tree-like structure, manual segmentation of the airway from chest CT scans is time-consuming and requires expert knowledge. Therefore, to relieve the burden of clinicians, automatic airway segmentation methods are proposed to extract airways accurately.

\begin{figure}[h]
\centering
\includegraphics[width=0.8\textwidth]{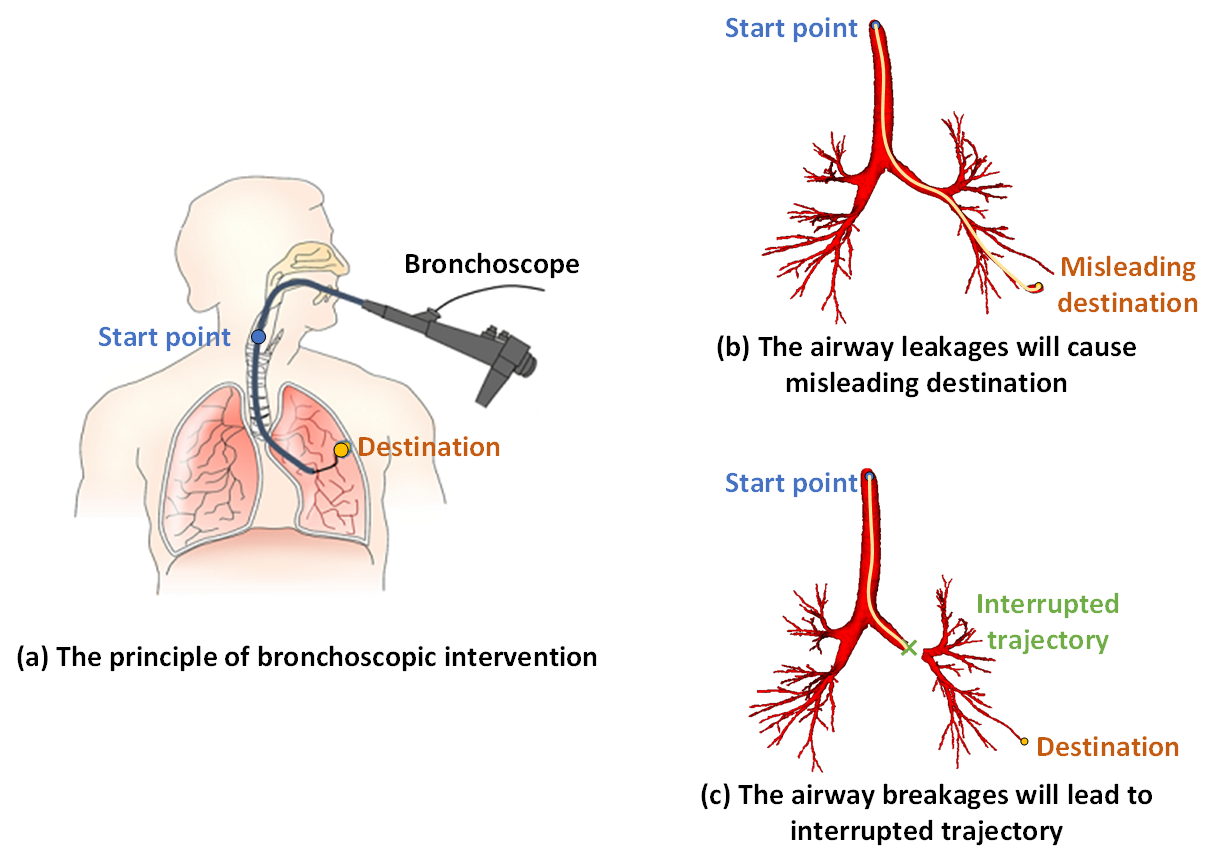}
\caption{The impact of leakages and breakages in the airway for bronchoscopic intervention. (a) describes the principle of bronchoscopic intervention referred to \cite{shen2019context}. (b) demonstrates that airway leakages will provide an incorrect destination and mislead the trajectory. (c) illustrates that airway breakages will interrupt the planning trajectory.}\label{Bronchoscopic intervention}
\end{figure}

Traditional airway segmentation methods are always designed based on region growing \cite{fabijanska2009two, chen2018automatic} and morphology operation \cite{aykac2003segmentation}, which are very sensitive to the manually designed features. Thus, their performance will be degraded once noises appear. Besides, owing to the similarity between bronchi and surrounding human tissues, these methods will cause severe leakages and breakages. As shown in Fig. \ref{Bronchoscopic intervention}, in bronchoscopic intervention, leakages in the airway will cause misleading destinations while breakages will lead to interrupted trajectories. Recently, convolutional neural networks(CNNs)-based methods \cite{juarez2018automatic, qin2020airwaynet, qin2020learning, zhang2021fda} are increasingly developed for airway extraction. U-Net \cite{ronneberger2015u} or 3D U-Net \cite{cciccek20163d} is widely used as the backbone to acquire a coarse airway segmentation result, which is then refined by utilizing prior knowledge like neighborhood connectivity \cite{qin2019airwaynet} or global information \cite{juarez2019joint}.

However, due to the complex tree-like structure of the airway, acquiring accurate and fine-grained segmentation results is difficult. The CNNs-based airway segmentation methods remain the following challenges. First, the intensity distribution is different in the main trachea region and bronchi region. In the main trachea region, the intensity contrast between the airway lumen and the wall is distinct and the features are easy to learn, while the ambiguity of intensity contrast in the bronchi region is hard to identify for the CNN models. Second, the encoder-decoder architecture is widely used in the CNN models like U-Net, where several pooling operations are utilized. Since some bronchi only have a diameter of 2-5 voxels, these features will be vanished by pooling operations and are hard to reconstruct in the decoder. Furthermore, these two challenges will result in weak airway features in the decoding process. Since adjacent slices in chest CT scans have similar airway shape and lumen position, one intuition for weak feature enhancement is to use this airway continuity prior knowledge. The prior knowledge could be achieved by applying slice feature propagation, so as to strengthen the weak feature and impose the continuity constraints.

To address the above problems and preserve the topological connectivity of the airway, we propose a two-stage end-to-end framework for accurate airway segmentation using long-term slice propagation (LTSP). In the proposed method, the slice relationship is considered in the decoding process. The slice feature is transferred in the proper direction to fully recover the continuity information destroyed in the encoder. Although slice propagation could alleviate the breakage phenomenon by enhancing the airway features, there remains a problem that slice features are hard to transfer among too long distances. In our method, to solve the above gradient vanish problem, we further design the LTSP cell in the decoding process where continuity information is extracted and densely propagates to other slices. Furthermore, we compare the proposed airway segmentation method to some state-of-the-art methods in 70 clinical CT scans. Extensive experiments show that our method achieves superior performance in extracting topological complete airway while maintaining the competitive overall segmentation accuracy.

% The main contributions in this paper are summarized as follows: (1) We propose an end-to-end framework to effectively segment accurate airways from chest CT scans. (2) We utilize the slice relationship in the decoding process to recover the lost airway information and enhance the connectivity of segmentation results. (3) We introduce dense architecture and long short-term memory mechanism to slice feature propagation and thus reduce the number of breakages.

\section{Methodology}\label{sec2}

\begin{figure}[h]
\centering
\includegraphics[width=1.0\textwidth]{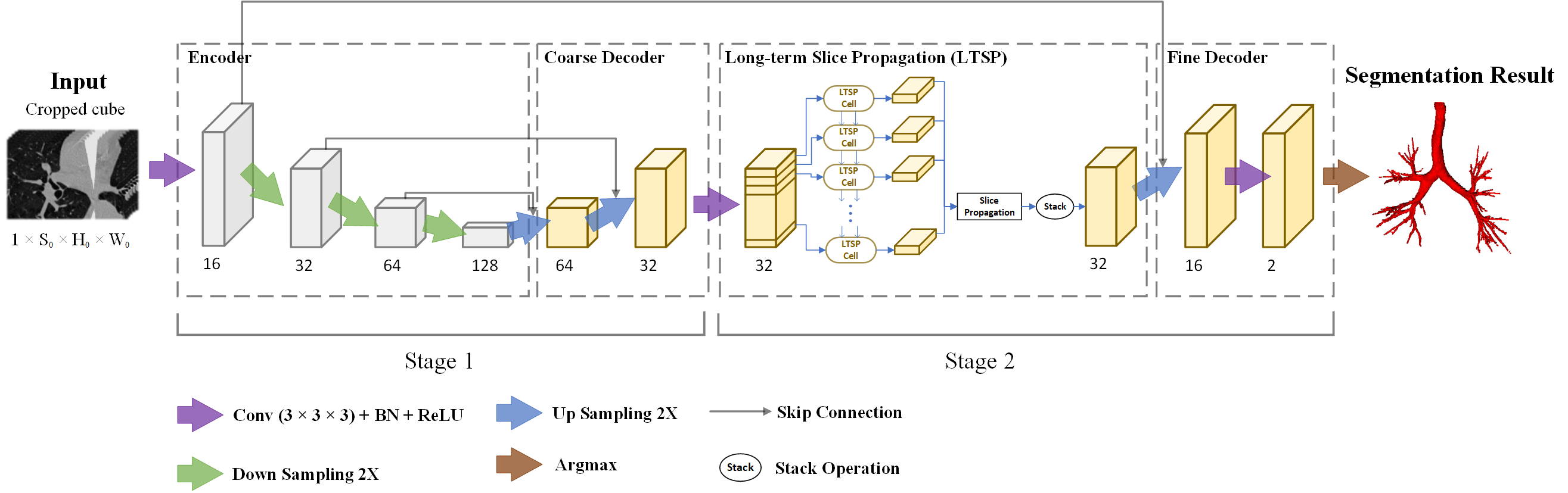}
\caption{Illustration of the proposed airway segmentation framework. The channel number is denoted below each feature map. In the first stage, a coarse feature map is extracted from 3D CT cropped cube by an encoder and a coarse decoder. In the second stage, each slice feature in the coarse feature map is propagated by passing through the LTSP cell. The results are stacked together and then used to predict the refined segmentation result by a fine decoder.}\label{Framework}
\end{figure}

The overview of the proposed airway segmentation method is illustrated in Fig. \ref{Framework}, which is a two-stage end-to-end framework. The first stage is used to acquire a coarse feature map by utilizing an encoder-decoder architecture. The second stage is to predict a refined and well-connected airway segmentation result by: (1) propagating slice information of the coarse feature map using the long-term slice propagation (LTSP) method, (2) decoding the airway segmentation result from the refined feature map with skip connection.

\subsection{Stage 1: Coarse feature map generation using encoder-decoder architecture}\label{sec2.1}

Stage 1 in the framework is to provide a coarse feature map by using an encoder and a coarse decoder. The feature map is extracted for subsequent slice propagation and airway refinement. 

In this stage, we employ an encoder-decoder architecture to extract the airway feature and then recover from the deep information. A part of 3D U-Net is used as the backbone, which contains three down-samplings and two up-samplings with skip connections. In each down-sampling module, two convolution layers (Conv) with batch normalization (BN) and rectified linear unit (ReLU) are followed by a max-pooling layer. In each up-sampling module, the up-sampled feature map and corresponding feature map in the encoder are concatenated together and then passed to a convolution layer with BN and ReLU.

Given a 3D CT cube input $X_0$ with size of $1 \times S_0 \times H_0 \times W_0$, a coarse feature map $X_{coarse}$ with size of $C \times S \times H \times W$ is generated by stage 1 as
\begin{equation}
    X_{coarse} = F_1(X_0),
	\label{eq1}
\end{equation}
where $F_1(\cdot)$ denotes the feature extraction process in stage 1.

\subsection{Stage 2: Segmentation refinement using long-term slice propagation}\label{sec2.2}

Stage 2 in the framework is used to densely propagate the slice features in the coarse feature map extracted in stage 1. The proposed LTSP method is designed to transfer the long-term slice features for the refined feature map generation. The final airway segmentation result is then predicted from the refined feature map by a fine decoder.

\subsubsection{Long-term slice propagation}\label{sec2.2.1}

Taking the airway's continuity prior knowledge into consideration, we utilize the slice propagation method into the decoder, which aims to enhance the weak airway features in each slice by integrating the adjacent slice features. Additionally, we improve the previous spatial CNN \cite{pan2018spatial} method and propose a long-term slice propagation (LTSP) method for effective feature transferring.

The original method proposed in \cite{pan2018spatial} uses the spatial CNN to propagate the neighboring message in 2D images. However, as it only transfers the feature from a slice to its nearest neighbor, the propagating airway feature will be rapidly dismissed and the continuity information will be lost. To improve the propagation efficiency and expand the receptive field of slice propagation, we propose the long-term slice propagation (LTSP) method. 

\begin{figure}[h]
\centering
\includegraphics[width=1.0\textwidth]{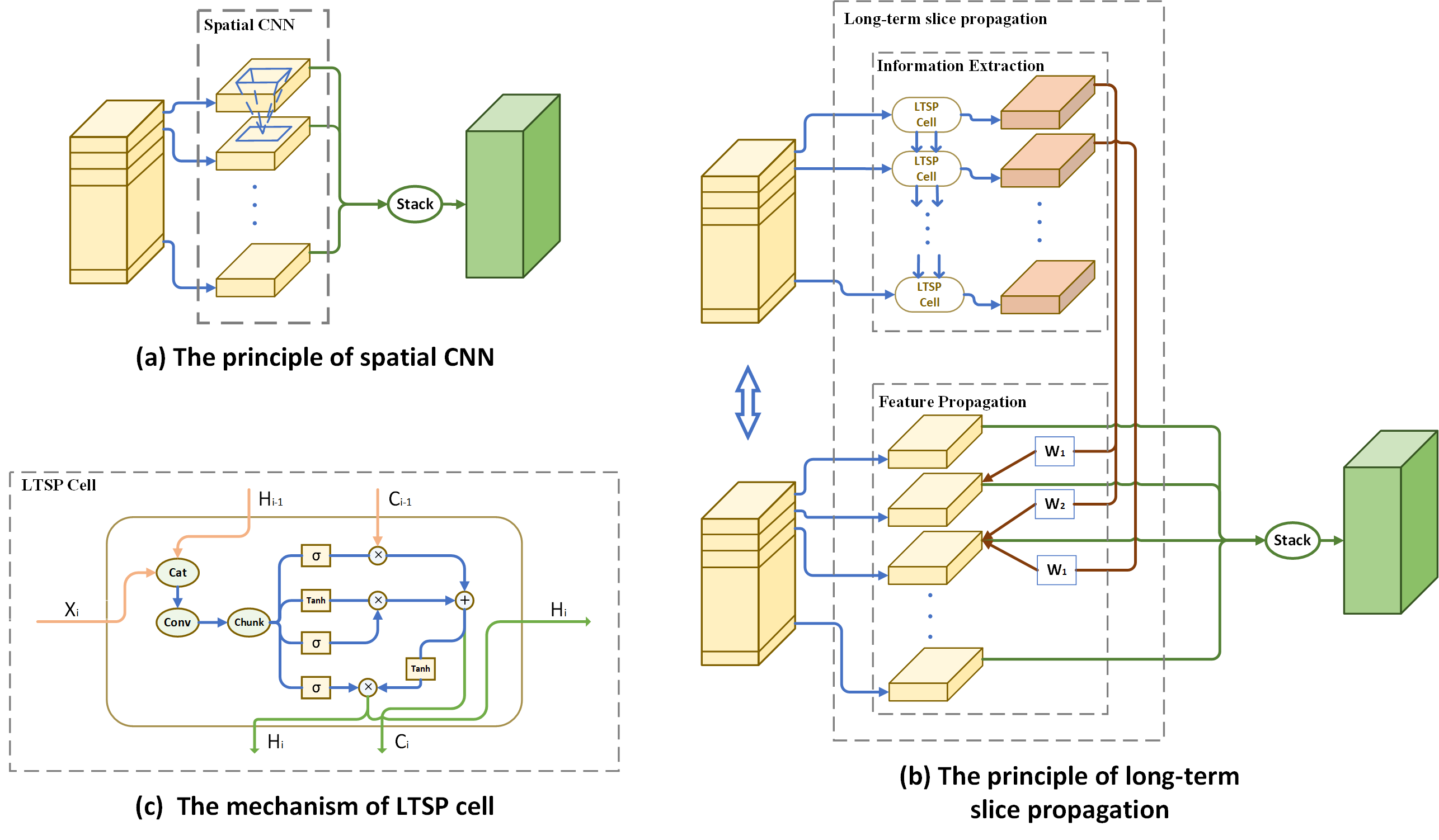}
\caption{Comparison of spatial CNN method and long-term slice propagation (LTSP) method. (a) gives the principle of spatial CNN where convolution operation is used to propagate the slice features in coarse feature map slice-by-slice. (b) illustrates the principle of long-term slice propagation. Each slice in the coarse feature map is passed through the LTSP cell to acquire a continuity information map which is then fused with the original feature by slice propagation. (c) demonstrates the mechanism of the designed LTSP cell. It receives the current slice feature $X_i$, previous output $H_{i-1}$ and previous cell state $C_{i-1}$ to acquire the current output $H_i$ and cell state $C_i$ by using the updating rules.}\label{Modules}
\end{figure}

% Given a coarse feature map $X_{coarse}$ with a size of $C \times S \times H \times W$, the LTSP method is achieved by the following steps: (1) Split the coarse feature map in the axial direction into $S$ slices. (2) Pass each slice into a designed LTSP cell to extract continuity information. (3) Integrate the extracted information with the next several slices' features. (4) Stack all propagated slices into a refined feature map.

Fig. \ref{Modules} gives the comparison of spatial CNN method and LTSP method. As illustrated in Fig. \ref{Modules} (a), in spatial CNN, one slice only receives the feature from its upper slice by directly utilizing convolution operation. For the complicated tree-like airway structure, there are several limitations in spatial CNN: (1) Insufficient feature propagation. Some breakages in the airway are dependent on the feature far from it, which needs to use features in several slices to recover it. As one slice only can receive the feature from its nearest neighbor, the information is not sufficient for its recovery. (2) Short propagation distance. Spatial CNN only uses convolution to extract the information from a neighbor slice, leading to the feature disappearance when the propagation distance is too long. Besides, a short propagation distance will result in a narrow receptive field and great propagation deficiency.

To address the problems, we propose the LTSP method to develop the effectiveness of slice propagation. As illustrated in Fig. \ref{Modules} (b), the LTSP method consists of two parts: information extraction and feature propagation.

In the information extraction part, the LTSP cell inspired by the long short-term memory (LSTM) \cite{hochreiter1997long} is designed to extract airway continuity information. Fig. \ref{Modules} (c) gives the propagating mechanism of the designed LTSP cell. Given a coarse feature map $X_{coarse}$ with the size of $C \times S \times H \times W$, each split slice $X_i$ with the size of $C \times H \times W$ is acquired by
\begin{equation}
\begin{aligned}
&X_i =  X_{coarse}[i].
\label{eq2}
\end{aligned}    
\end{equation}
Together with the previous LTSP cell output $H_{i-1}$ and previous cell state $C_{i-1}$ (with each size of $C \times H \times W$), the forward calculation includes the following three steps. First, the stacked gates $G$ is acquired by using 2D convolution operation to double the channel number of the concatenation of $X_i$ and $H_{i-1}$, where $G$ is computed by
\begin{equation}
\begin{aligned}
&G =  Conv(Cat(X_i, H_{i-1})).
\label{eq3}
\end{aligned}    
\end{equation} Then, the convolution result $G$ is chunked into $G_1$, $G_2$, $G_3$ and $G_4$, with each size of $C \times H \times W$. Pass them into different activation function to calculate forget gate $G_f$, new cell state $\tilde{C_i}$, input gate $G_i$, and output gate $G_o$, respectively. The calculation process could be described as
\begin{equation}
\begin{aligned}
&G_f = \sigma(G_1), \qquad\quad
\tilde{C_i} = \tanh(G_2), \\
&G_i = \sigma(G_3), \qquad\quad\
G_o = \sigma(G_4). 
\label{eq4}
\end{aligned}    
\end{equation} Finally, previous cell state $C_{i-1}$ and new cell state $\tilde{C_i}$ are used to calculate the updated output $H_{i}$ and final cell state $C_{i}$ by 
\begin{equation}
\begin{aligned}
&C_i = G_f \circ C_{i-1} + G_i \circ \tilde{C_{i}}, \\
&H_i = G_o \circ tanh(C_i).
\label{eq5}
\end{aligned}    
\end{equation}

In the feature propagation part, the original slice $X_i$ is updated by propagating information maps extracted by LTSP cells. The forward slice propagation could be described as 

\begin{equation}
    X^{'}_{i} = 
	\begin{cases}
	X_i, \qquad\qquad\qquad\qquad\qquad\qquad\qquad\qquad\quad\; i = 1,\\
	
	X_i + f\big({H_{i-1} * W_1}), \qquad\qquad\qquad\qquad\qquad i = 2,\\
	
	X_i + f\big(H_{i-1} * W_1 + H_{i-2} * W_2), \qquad\qquad\  {i = 3, ..., S},
	\end{cases}
	\label{eq6}
\end{equation}
where $W_1$ and $W_2$ denote the convolution weights for 1-distance feature and 2-distance feature, respectively. And $f(\cdot)$ denotes the nonlinear activation function like ReLU. 

In the end, all updated slices $X^{'}_i$ are stacked into a new refined feature map $X_{fine}$ with the size of $C \times S \times H \times W$ by
\begin{equation}
    X_{fine} = [X_1^{'}, X_2^{'}, ... , X_S^{'}].
    \label{eq7} 
\end{equation}

\subsubsection{Segmentation prediction and optimization}\label{sec2.2.2}

To acquire the airway segmentation from refined feature map $X_{fine}$, a fine decoder with skip connection is used to get the probability map of the airway $X_{prob}$. An argmax operation is then utilized to predict the final airway segmentation result $X_{seg}$. The prediction process could be described as
\begin{equation}
\begin{aligned}
&X_{prob} = F_2(X_{fine}), \\
&X_{seg} = argmax(X_{prob}),
\label{eq8}
\end{aligned}    
\end{equation}
where $F_2(\cdot)$ denotes the up-sample operation in fine decoder.

In the training process, soft dice loss \cite{milletari2016v} is used for the airway segmentation tasks. Given prediction $p(x)$ and corresponding binary label $y(x)$ for each voxel $x$ in segmentation result volume $X$, the segmentation loss could be calculated by

\begin{equation}
\begin{aligned}
\mathcal{L}_{seg} = 1 - \frac{2\sum_{x\in X}p(x)y(x) + \epsilon}{\sum_{x\in X}(p(x) + y(x)) + \epsilon},
\label{eq9}
\end{aligned}    
\end{equation}
where smoothing parameter $\epsilon$ is used to avoid division by zero.

\section{Experiments and results}\label{sec3}

We evaluate the proposed method in LIDC dataset \cite{qin2019airwaynet}, where 50 chest CT scans are randomly chosen for training and the remaining 20 scans are used for testing. Furthermore, ablation studies are conducted to confirm the effectiveness of our method.

\subsection{Datasets and implementation details}\label{sec3.1}

The experiment dataset contains 70 clinical chest CT scans, where the pixel spatial resolution ranges from 0.5 to 0.781mm and slice thickness varies from 0.45 to 1.0mm. The model is trained on 50 randomly chosen scans and tested on the remaining 20 scans.

To improve the model's generalization ability, the HU value of each scan is truncated into [-1000, 600] and linearly mapped into [0, 1]. Besides, data augmentation is performed to each cropped cube including horizontal flipping and slight rotation. For training strategy, we adopt Adam optimizer \cite{kingma2014adam} ($\beta_1 = 0.9$, $\beta_2 = 0.999$) with learning rate set as 0.002. Our method is implemented in PyTorch 1.7 with NVIDIA GeForce RTX 3090. The training process converges in 30 epochs.

Additionally, we utilize the center crop method instead of the random crop method to acquire more representative cropped cubes as the model's inputs. The center crop method is used to guarantee that the cropped cubes are able to contain sufficient airway features. Given a CT scan with a size of $S^{'} \times H^{'} \times W^{'}$, we first search for the minimum and maximum index for airway region in x, y, and z directions. We denote these index pairs as $(X_{min}, X_{max}), (Y_{min}, Y_{max}), (Z_{min}, Z_{max})$ and randomly choose a voxel $P = (x, y, z)$ from these spans as the cropped center. The input cube and corresponding binary label with a size of $S_0 \times H_0 \times W_0$ are cropped from original CT scans and ground truth maps based on the cropped center. Then the cropped cubes are fed into the model for training. Furthermore, we utilize the sliding window technique in the testing process to acquire each cube's prediction and combine them to form the whole airway segmentation result.

\subsection{Evaluation metrics and results}\label{sec3.2}

To evaluate our method, we adopt three metrics to assess the topological completeness and segmentation accuracy of airway prediction results: (1) Branches detected (BD), (2) Tree-length detected (TD), and (3) Dice coefficient (DSC). The definition of three metrics could be referred to \cite{lo2012extraction}.

In our method, 3D U-Net is adopted to be the backbone of the segmentation framework. Therefore, we compare our method with original 3D U-Net \cite{cciccek20163d} and other related state-of-the-art methods like Wang et al. \cite{wang2019tubular} and Juarez et al. \cite{juarez2019joint}. Wang et al. \cite{wang2019tubular} develop a radial distance loss for detecting more tiny airway tubular structures. Juarez et al. \cite{juarez2019joint} introduce graph neural network (GNN) module into the deepest level of 3D U-Net to improve airway connectivity. These methods are implemented by ourselves and fine-tuned on the experiment dataset. Table \ref{tab1} gives the comparison results which show that our method achieves the best performance in BD and TD with compelling DSC. Compared to others, our method increases the BD and TD by over 3 \% and 2 \%, respectively. Since these two metrics could reflect the topological completeness of airway segmentation, the comparison results also demonstrate that our method outperforms the others in detecting more small bronchi and improving the connectivity of airway segmentation.

\begin{table}[ht]
\begin{center}
\begin{minipage}{260pt}
\caption{Results (\%) of the proposed framework compared to state-of-the-art methods (Mean$\pm$Standard deviation).}\label{tab1}
\begin{tabular}{@{}llll@{}}
\toprule
Method   & BD       & TD      & DSC \\
\midrule
3D U-Net \cite{cciccek20163d} & $86.06\pm11.72$   & $83.10\pm9.62$  & $93.35\pm1.74$  \\
Wang et al. \cite{wang2019tubular}     & $87.97\pm9.49$   & $84.61\pm8.54$  & $93.12\pm2.00$  \\
Juarez et al. \cite{juarez2019joint}     & $87.40\pm10.15$   & $84.60\pm8.94$  & $\mathbf{93.51\pm1.57}$  \\
Our proposed    & $\mathbf{90.83\pm9.38}$   & $\mathbf{87.59\pm8.71}$  & $92.95\pm1.61$  \\
\botrule
\end{tabular}
\end{minipage}
\end{center}
\end{table}

\begin{figure}[h]
\centering
\includegraphics[width=1.0\textwidth]{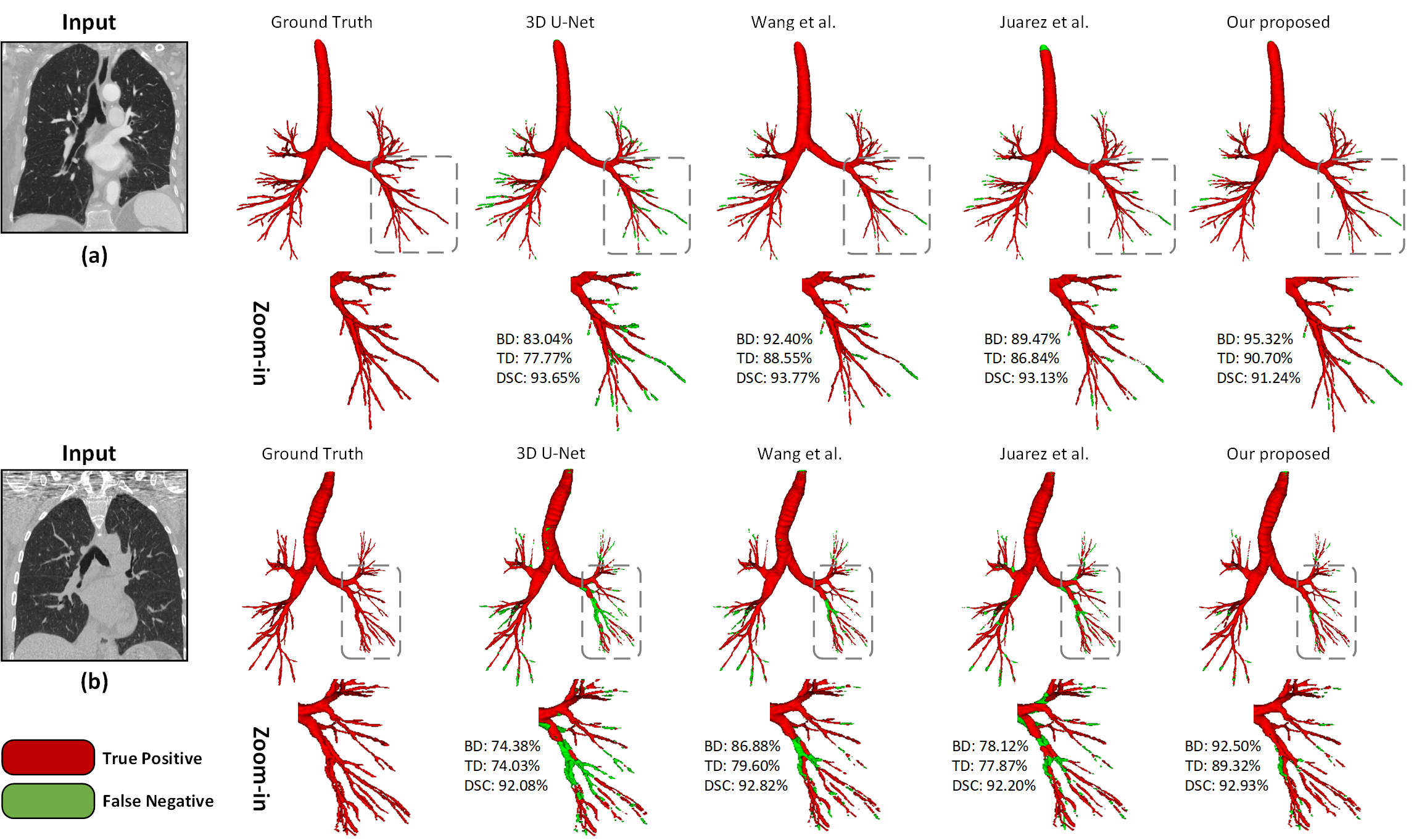}
\caption{Rendering of airway segmentation results. (a) and (b) give the comparison of different methods in an easy case and a hard case, respectively. The true positive voxels are shown in red color, while the false negative voxels are shown in green color.}\label{Qualitative Results}
\end{figure}

Qualitative comparison of airway segmentation shown in Fig. \ref{Qualitative Results} also demonstrates the effectiveness of our method. Compared to other methods, more small branches and tiny bronchi are reconstructed in our method, which results in the great improvement of the airway topological connectivity. The preservation of airway topological completeness is significant for clinical practice like bronchoscopic intervention and surgical navigation. In Fig. \ref{Qualitative Results} (b), a great number of breakages (false negatives voxels in green color) are eliminated in our method while maintaining a high overall segmentation accuracy. 

% Originally, these breakages are produced due to the weak and insufficient airway features in each slice, which could be enhanced by performing slice feature propagation in our method. 

\subsection{Ablation study}\label{sec3.3}

We also conduct ablation studies to further investigate the effect of each component in our method. Since we use dense structures in the LTSP method, the density of this structure should be considered in our framework. Therefore, experiment comparisons are made by utilizing modules with different propagation distances in a single pass. Furthermore, to validate the effectiveness of the LTSP cell in slice propagation, experiments with or without LTSP cells are also conducted. 

\begin{table}[ht]
\begin{center}
\begin{minipage}{305pt}
\caption{Results (\%) of ablation study on the testing set (Mean$\pm$Standard deviation).}\label{tab2}
\begin{tabular}{@{}llll@{}}
\toprule
Propagation distance   & BD       & TD      & DSC \\
\midrule
No propagation module & $86.06\pm11.72$   & $83.10\pm9.62$  & $\mathbf{93.35\pm1.74}$  \\
One-slice module     & $87.97\pm9.49$   & $84.61\pm8.54$  & $93.12\pm2.00$  \\
One-slice module + LTSP cells     & $90.11\pm10.28$   & $87.20\pm8.96$  & $93.15\pm1.79$  \\
Two-slices module     & $89.22\pm9.41$   & $86.39\pm7.84$  & $92.42\pm2.79$  \\
Two-slices module + LTSP cells     & $\mathbf{90.83\pm9.38}$   & $\mathbf{87.59\pm8.71}$  & $92.95\pm1.61$  \\

\botrule
\end{tabular}
\end{minipage}
\end{center}
\end{table}

As shown in Table \ref{tab2}, the LTSP cells and more densely-connected structure both can achieve higher BD and TD with DSC maintained. For the comparison of propagation distance, the two-slices module outperforms the no propagation module and one-slice module since it enables the slice feature to propagate to more slices directly. Furthermore, experiment results show that the designed LTSP cell could effectively improve the segmentation accuracy and topological completeness by extracting more valuable continuity information. 
% Besides, considering the propagation direction, we summarize from the experiments that propagating from top to bottom could achieve the best performance, which corresponds to the fact that the overall airway grows along the axial direction.

\section{Conclusion}\label{sec4}

This paper proposed an effective long-term slice propagation (LTSP) method for accurate airway segmentation. Focusing on alleviating breakage phenomenon and improving topological connectivity, LTSP cell and dense slice propagation were designed to fully exploit the slice continuity relationship. Extensive experiments showed that the proposed method outperformed some state-of-the-art methods by detecting more tiny bronchi and reconstructing essential branches, which further validates the effectiveness of its constituents. The proposed method is beneficial to extracting topological complete airway segmentation for bronchoscopic intervention.

%%===========================================================================================%%
%% If you are submitting to one of the Nature Portfolio journals, using the eJP submission   %%
%% system, please include the references within the manuscript file itself. You may do this  %%
%% by copying the reference list from your .bbl file, paste it into the main manuscript .tex %%
%% file, and delete the associated \verb+\bibliography+ commands.                            %%
%%===========================================================================================%%

\bibliography{sn-bibliography}% common bib file
%% if required, the content of .bbl file can be included here once bbl is generated
%%\input sn-article.bbl

%% Default %%
%%\input sn-sample-bib.tex%

\end{document}